\documentclass[aps,prd,twocolumn,superscriptaddress,nofootinbib]{revtex4-2}

\usepackage[colorlinks=true,citecolor=red,urlcolor=blue]{hyperref}
\usepackage{amssymb}
\usepackage{amsmath,color}
\usepackage{graphicx}
\usepackage{bbm}
\usepackage{verbatim}
\usepackage[T1]{fontenc}
\usepackage[utf8]{inputenc}
\usepackage[normalem]{ulem}

\usepackage{url}
\usepackage{xcolor}
\usepackage{caption}
\usepackage{subcaption}

\usepackage{diagbox}
\usepackage{color}
\usepackage{multirow}
\usepackage{hhline}
\usepackage{tabularx}

\newcommand{\orcid}[1]{\href{https://orcid.org/#1}{\hspace{0.5mm}\raisebox{-0.ex}{\includegraphics[height=2.0ex]{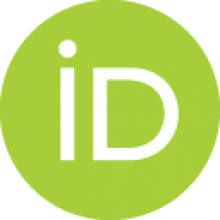}}}}

\begin{document}

\title{Eccentricity enables the earliest warning and localization of gravitational waves with ground-based detectors}

\author{Tao Yang\orcid{0000-0002-2161-0495}}
\affiliation{School of Physics and Technology, Wuhan University, Wuhan 430072, China}
\affiliation{Center for the Gravitational-Wave Universe, Astronomy Research Center, Seoul National University, 1 Gwanak-ro, Gwanak-gu, Seoul 08826, Korea}

\author{Rong-Gen Cai\orcid{0000-0002-3539-7103}} 
\affiliation{CAS Key Laboratory of Theoretical Physics, Institute of Theoretical Physics, Chinese Academy of Sciences, Beijing 100190, China}
\affiliation{School of Physical Science and Technology, Ningbo University, Ningbo, 315211, China}
\affiliation{School of Fundamental Physics and Mathematical Sciences, Hangzhou Institute for Advanced Study (HIAS), University of Chinese Academy of Sciences, Hangzhou 310024, China}

\author{Zhoujian Cao\orcid{0000-0002-1932-7295}}
\email[Corresponding author: ]{zjcao@bnu.edu.cn}
\affiliation{Department of Astronomy, Beijing Normal University, Beijing 100875, China}
\affiliation{School of Fundamental Physics and Mathematical Sciences, Hangzhou Institute for Advanced Study (HIAS), University of Chinese Academy of Sciences, Hangzhou 310024, China}

\author{Hyung Mok Lee\orcid{0000-0003-4412-7161}}
\affiliation{Center for the Gravitational-Wave Universe, Astronomy Research Center, Seoul National University, 1 Gwanak-ro, Gwanak-gu, Seoul 08826, Korea}

\date{\today}

\begin{abstract}

The early and precise localization of gravitational waves (GWs) is pivotal in detecting their electromagnetic (EM) counterparts, 
especially for binary neutron stars (BNS) and neutron star-black hole binaries (NSBH).
In this letter, we pioneer the exploration of utilizing the higher harmonic modes induced by the eccentricity of compact binaries to localize GWs 
with ground-based detectors even before the quadrupole baseline $\ell=2$ mode enters the detector band.
Our theoretical analysis marks a first in proposing a strategy for gaining the earliest possible warning and maximizing preparation time 
for observing pre- and/or post-merger EM counterparts.
We simulate three typical binaries from GWTC-3 with eccentricities ranging from 0.05 to 0.4. 
Our results reveal that the third-generation (3G) detectors (low frequency cut-off $f_0=5$ Hz)
can accumulate sufficient signal-to-noise ratios through higher modes 
before the onset of the baseline $\ell=2$ mode entry into the band. 
Notably, relying solely on the higher modes, the 3G detector network ET+2CE achieves an average localization 
on the order of $1-10^2~\rm deg^2$ around 1-1.8 hours before the merger of a GW170817-like BNS,
and $10-10^3~\rm deg^2$ approximately 18-30 minutes prior to the merger of a GW200115-like NSBH.
A $100~\rm deg^2$ localization is attainable even 2-4 hours prior to a BNS merger.
Moreover, in the near face-on orientations which are generally more favorable for EM counterpart detection, the localization can be further improved.
\end{abstract}

\maketitle

\section{Introduction}
Multi-messenger observations of gravitational waves (GWs) and their electromagnetic (EM) counterparts play vital roles in cosmology, astrophysics, and fundamental physics
\citep{LIGOScientific:2017ync,LIGOScientific:2017zic,LIGOScientific:2017adf,Creminelli:2017sry,Ezquiaga:2017ekz,Baker:2017hug,Mooley:2018qfh,Hotokezaka:2018dfi,Dietrich:2020efo}. 
Notably, with the observations of EM counterparts, the host galaxies of GWs and consequently their redshifts can be readily identified, 
enabling the direct measurement of the Hubble constant through GW standard sirens~\citep{Holz:2005df,Dalal:2006qt,Nissanke:2009kt,LIGOScientific:2017adf}
(for dark sirens with no EM counterparts see e.g.
\citep{Chen:2017rfc,Borhanian:2020vyr,LIGOScientific:2018gmd,DES:2019ccw,Gray:2019ksv,DES:2020nay,Borhanian:2020vyr,Finke:2021aom}).
However, the successful capture of EM counterparts heavily relies on the precise and timely localization of GW sources
\citep{Petrov:2021bqm,Chu:2015jxa,Ghosh:2015sxp,Coughlin:2018lta}.
The early warning and localization of GWs, as well as their implications for multi-messenger observations, have been explored from various perspectives
\citep{Cannon:2011vi,Kyutoku:2013mwa,Chan:2018csa,Nitz:2020vym,Singh:2020lwx,Kapadia:2020kss,Singh:2022tlh,Tsutsui:2020bem,Magee:2021xdx,Magee:2022kkc,Tsutsui:2021izf,Chatterjee:2022dik,Li:2021mbo,Miller:2023rnn,Sachdev:2020lfd,Kovalam:2021bgg,Chaudhary:2023vec,Hu:2023hos}.

Eccentricity can aid not only in distinguishing between isolated and dynamical binary black hole (BBH) formation scenarios 
\citep{Nishizawa:2016jji,Nishizawa:2016eza,Breivik:2016ddj,Zevin:2021rtf} 
but also in improving the parameter estimation (including localization) of GWs~\citep{Sun:2015bva,Ma:2017bux,Pan:2019anf,Mikoczi:2012qy}.
In particular, recent studies~\citep{Yang:2022tig,Yang:2022fgp,Yang:2022iwn}
demonstrate that the eccentricity of long inspiraling compact binaries can dramatically enhance the accuracy of distance estimation
and source localization by several orders of magnitude with space-based decihertz observatories. 
On the other hand, neglecting eccentricity can introduce biases in parameter estimation and in testing general relativity with GWs
\citep{Favata:2013rwa,Favata:2021vhw,GilChoi:2022nhs,Saini:2022igm,Narayan:2023vhm}. 
In current detections, GW190521 has been suggested to favor nonzero eccentricities, 
and its effects on parameter estimation have been investigated from various perspectives
\citep{Romero-Shaw:2020thy,Gayathri:2020coq,Gayathri:2020mra,LIGOScientific:2020ufj,LIGOScientific:2020iuh}.
These research findings suggest that eccentricity is an indispensable factor to consider in GW detection, data analysis, and practical applications.

One prominent feature of eccentric waveforms is the presence of multiple harmonic modes induced by eccentricity \citep{Yunes:2009yz,Huerta:2014eca,Moore:2018kvz,Moore:2019xkm}. 
In the quasi-circular case, quadrupole GWs exhibit only the baseline mode, $\ell=2$, 
whose frequency is twice of the orbital frequency $F$ (hereafter, we refer to the $\ell=2$ mode as the baseline mode).
Eccentricity can induce higher harmonic modes, each with a frequency $f_{\ell}(t)=\ell F(t)$.
This results in each mode entering the detector band at a distinct time. 
Consequently, the higher modes ($\ell>2$) enter the detector band earlier than the dominant baseline mode,
affording an extended period for the observation of these higher modes.
This is particularly beneficial for ground-based detectors that aim to capture GWs at high frequencies, where the inspiral time is very limited.
For instance, consider the binary neutron stars (BNS) GW170817 detected by LIGO as an example.
The in-band time of the baseline mode from 10 Hz to the merger lasts approximately 17 minutes.
However, before the baseline mode enters the detector band, the higher modes have already been in band for nearly 20 hours
(assuming we account for the higher modes up to $\ell=10$; note that a larger eccentricity would shorten this period).
This extended observation window for the higher modes enables the consideration of Earth's rotation effects, 
introducing a Doppler effect that can furnish additional angular information regarding the sources.
A nonvanishing eccentricity can render the contribution of these higher modes non-negligible even before the baseline mode becomes observable. 
The early detection and localization of these higher modes on their own have not been considered in previous works.
This raises the following questions that we aim to address in this letter:
Can we observe these higher modes even before the baseline mode becomes observable?
More importantly, 
to what degree of accuracy can we localize GW sources solely based on these higher modes in the very early stage?

Previous works on the early warning of GWs have focused on the time following the entry of the dominant baseline mode into the band
\cite{Kyutoku:2013mwa,Kapadia:2020kss,Singh:2022tlh}. 
They aim at comparing the circular case (only baseline mode) to the eccentric case (baseline + higher modes)~\cite{Kyutoku:2013mwa},
or the quadrupole mode to the quadrupole + higher multipoles~\cite{Kapadia:2020kss,Singh:2022tlh}.
In this letter, for the first time, we shift our focus to the time preceding the baseline dominant mode's entry. 
This means we exclusively rely on the higher modes in the very early stage.
Localization based on these higher modes allows for the earliest detection and warning of GWs and electromagnetic (EM) counterparts, 
especially in the context of BNS and neutron star-black hole (NSBH) binaries. 
This grants EM telescopes much more preparation time, 
enhancing the likelihood of capturing EM counterparts and particularly aiding in the capture of potential pre-merger EM counterparts
\citep{Tsang:2011ad,Suvorov:2020tmk,Lyutikov:2018nti,Paschalidis:2013jsa}.
If there is a high likelihood that most NSBH mergers will involve non-disruptive systems, 
pre-merger signals might provide the sole avenue for EM observations of these systems.

\section{Methodology}
To address the aforementioned questions, 
we turn to simulations of typical compact binaries that have detected by LIGO-Virgo-KAGRA collaborations. 
We select three representative binaries from the GWTC-3 catalog~\citep{LIGOScientific:2021djp}:
BNS GW170817, NSBH GW200115, and binary black holes (BBH) GW150914.
We consider various detector network scenarios, 
including the second-generation (2G) advanced LIGO, advanced Virgo, KAGRA, and India-LIGO (HLVKI) networks at their designed sensitivity, 
the third-generation (3G) detector Einstein Telescope (ET)~\citep{Punturo:2010zz}, 
and the extended network ET+2CE, which combines ET with two cosmic explorer detectors~\footnote{\url{https://gwic.ligo.org/}}. 
The low-frequency cutoffs for the 2G and 3G detector bands are set at 10 Hz and 5 Hz, respectively.
It's worth noting that the lower limit of ET can reach 3 Hz~\footnote{\url{https://apps.et-gw.eu/tds/?content=3&r=17196}}, 
which potentially yields more promising results.
In this study, we conservatively assume a uniform lower limit of 5 Hz for all 3G detectors.

We adopt the non-spinning, inspiral-only, and frequency-domain EccentricFD waveform
approximant provided by the {\sc LALSuite} software package~\citep{lalsuite}. 
This waveform is sufficient for our purposes, as we focus on the early stage of the inspiral phase.
The eccentric waveforms are generated using {\sc PyCBC}~\citep{alex_nitz_2022_6912865}.  
The waveform can be expressed as the sum of  harmonics~\citep{Huerta:2014eca}: 
\begin{equation}
\tilde{h}(f) = \sum_{\ell=1}^{10} \tilde{h}_\ell(f)\,, 
\label{eq:hf}
\end{equation}
where $\tilde{h}_\ell(f)$ represents the contribution from the $\ell$-th harmonic.
We modify {\sc LALSuite} to extract each harmonic $\tilde{h}_\ell(f)$ separately.

Given the relationship $f_\ell(t)=\ell F(t)$, the time associated with a fixed frequency varies for each mode. 
Accounting for Earth's rotation, the antenna pattern functions $F_{+,\times}$ of the detector become time-dependent.
Consequently, it's crucial to handle the antenna pattern function for each mode individually, ensuring accurate representation of their time-varying behavior.
To establish the relationship between time and frequency $t(f)$ for eccentric binaries, 
we use the baseline frequency $f_2$ as a reference and numerically solve the phase evolution of the eccentric orbits~\citep{Yunes:2009yz}.
Then for each mode $\tilde{h}_\ell(f)$, the associated response functions are $F_{+,\times}(t)$, with the time $t(f\to 2f/\ell)$.
To assess the contribution of higher modes at a particular time, such as during the period before the baseline mode enters the band,
each mode in Eq.~(\ref{eq:hf}) must be truncated at its respective frequency, which corresponds to the time when the baseline mode starts to enter the band.

We adopt the approach of~\citep{Yang:2022tig} and employ the Fisher matrix technique for GWs~\citep{Cutler:1994ys} 
to estimate the uncertainty and covariance of the waveform parameters. 
The Fisher matrix is defined as $\Gamma_{ij}=(\partial_i h,\partial_j h)$, 
where $\partial_i h=\partial h/\partial P_i$ and $P_i$ represents a parameter in the waveform
(refer to~\citep{Yang:2022tig,Yang:2022fgp} for details). 
The inner product is defined as 
\begin{equation}
(a,b)=4\int_{f_{\rm min}}^{f_{\rm max}}\frac{\tilde{a}^*(f)\tilde{b}(f)+\tilde{b}^*(f)\tilde{a}(f)}{2 S_n(f)}df\,,
\label{eq:inner}
\end{equation}
here $f_{\rm min}$ is the low-frequency cutoff of the detector, $S_n(f)$ is the noise power spectral density. Then the signal-to-noise ratio (SNR) is $\rho=(h,h)$.
The sky localization error is given by $\Delta \Omega=2\pi |\sin(\theta)|\sqrt{C_{\theta\theta}C_{\phi\phi}-C_{\theta\phi}^2}$~\citep{Cutler:1997ta}, 
with the covariance matrix of the parameters as $C_{ij}=(\Gamma^{-1})_{ij}$. 
We need to note the limitations of the Fisher matrix in the parameter estimation of GWs, especially for low SNRs~\cite{Vallisneri:2007ev}.
We incorporate Gaussian priors $\Gamma^p_{ii}=1/(\delta P_i)^2$ into the Fisher information matrix,
where $\delta P_i$ represents the maximum permissible variation in the parameter~\citep{Cutler:1994ys,Favata:2013rwa,Favata:2021vhw,Cho:2022cdy}.
The Fisher matrix approach, with the incorporation of Gaussian priors, 
has demonstrated consistency with the more computationally intensive Markov Chain Monte Carlo (MCMC) method
\citep{Favata:2021vhw,Cho:2022cdy}. 

To account for the different locations of the detectors and the effects of Earth's rotation, 
we use a geocentric coordinate system to derive the time-dependent antenna pattern functions of the detectors. 
Then, based on the relation $t(f)$ for each harmonic mode, we finally obtain the frequency-dependent antenna pattern functions 
(refer to~\citep{Yang:2022fgp,Graham:2017lmg,Yang:2022iwn} for the similar technical details).
The positions and orientations of the arms for the 2G detector network HLVKI are detailed in Table I of~\citep{Pan:2019anf}.
However, the exact locations and orientations of the 3G detectors ET and CE have not been finalized. 
In our study, we assume ET is situated in the Netherlands, 
while the two CE detectors share the same locations and orientations as LIGO Hanford and Livingston
\footnote{The exact detector location and orientation are not critical for this study and will not affect our results, as we simulate binaries uniformly distributed across the sky, aiming for average results.}. 

For each typical binary, we maintain the component masses ($m_1,~m_2$), redshift $z$, 
and luminosity distance $d_L$ consistent with the median values of the actual events~\footnote{\url{https://gwosc.org/eventapi/html/GWTC/}}.
For angular parameters $P_{\rm ang}$, such as the inclination angle $\iota$, sky location ($\theta, \phi$), polarization angle $\psi$, 
and longitude of ascending nodes axis $\beta$, 
we generate 1000 random sets from a uniform and isotropic distribution.
Given the validated use of this eccentric waveform for initial eccentricities up to 0.4~\citep{Huerta:2014eca}, 
we consider four discrete initial eccentricities: $e_0=0.05,~0.1,~0.2, \text{ and } 0.4$. 
It's important to note that $e_0$ is defined at $f_0=10$ Hz for 2G detectors and $f_0=5$ Hz for 3G detectors. 
Given that eccentricity is roughly inversely proportional to frequency~\citep{Yunes:2009yz}, 
the same eccentricity value in the 3G detector context is more conservative than in the 2G detector scenario.
Without loss of generality, we set the coalescence time and phase as $t_c=\phi_c=0$. 

\section{Results}
\begin{figure*}
\centering
\includegraphics[width=0.9\textwidth]{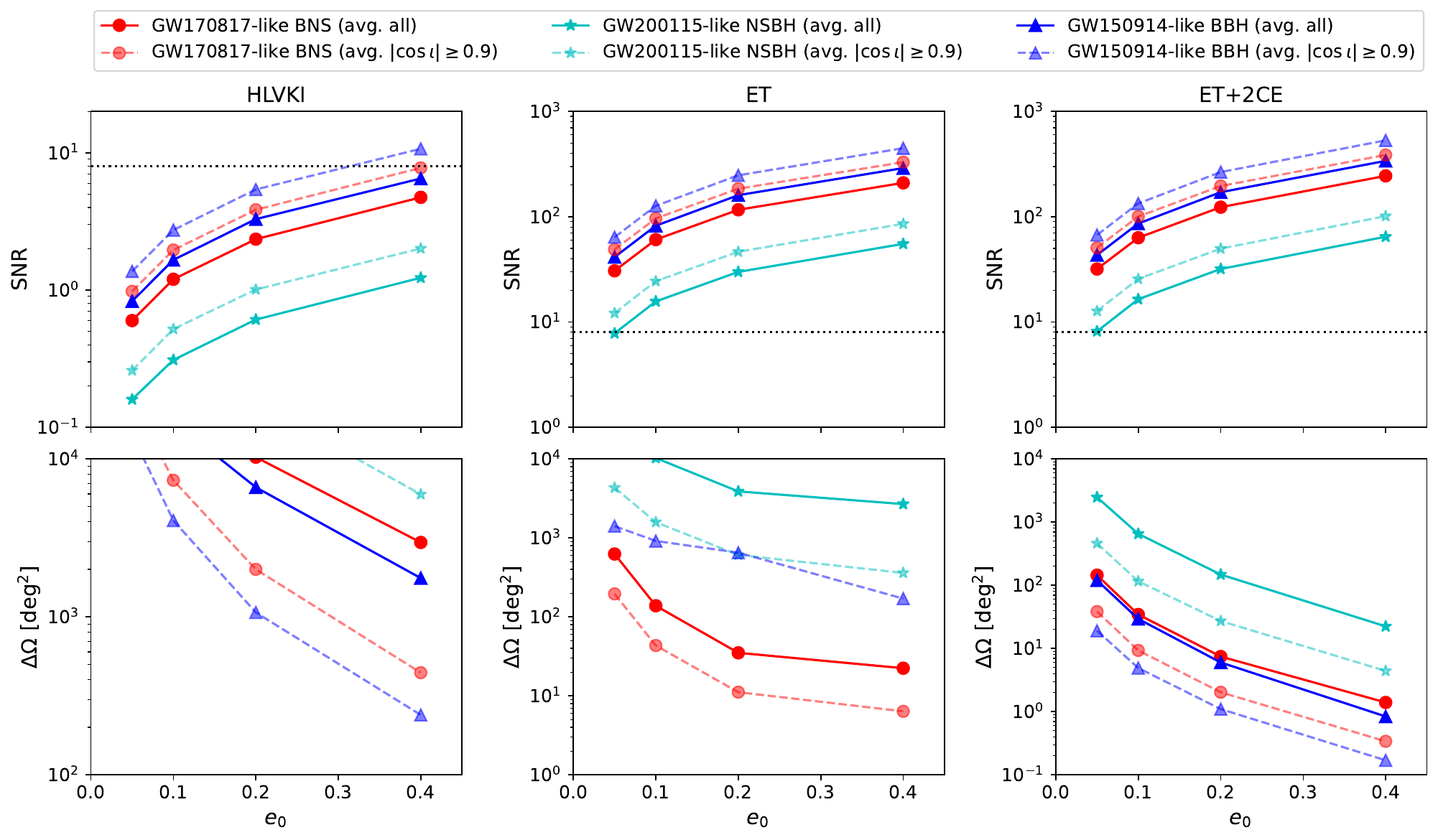}
\caption{The average SNR and localization provided exclusively by the higher harmonic modes just before the baseline mode enters the band.
The luminosity distances for BNS, NSBH, and BBH are 40, 290, and 440 Mpc, respectively.
The solid line represents the average of all 1000 $P_{\rm ang}$ samples, 
while the dashed line corresponds to the average with $|\cos\iota|\ge 0.9$ (near face-on orientations). 
The dotted lines in the upper panel indicate the threshold SNR, $\rho_{\rm th}=8$. 
Only cases with $\Delta\Omega<10^4~\rm deg^2$ are shown.}
\label{fig:snr_loc_pre}
\end{figure*}

Fig.~\ref{fig:snr_loc_pre} shows the average SNR and localization, 
provided exclusively by the higher modes ($\ell>2$), just before the dominant baseline mode ($\ell=2$) enters the band.
We utilize two averaging strategies: first, by averaging across all 1000 $P_{\rm ang}$ samples, 
and second, by averaging the $P_{\rm ang}$ samples that meet the condition $|\cos\iota|\ge 0.9$.
We define the condition $|\cos\iota|\ge 0.9$ (equivalently, $|\iota|\leq 25^\circ$ or $|180^{\circ}-\iota|\leq 25^\circ$) as near face-on orientations. 
These orientations are optimal for achieving both a higher SNR and enhanced localization of the GWs.
Furthermore, if we detect the EM counterparts of BNS and NSBH, 
they are more likely to be observed in near face-on orientations~\citep{Fermi-LAT:2009owx,Nakar:2005bs,Rezzolla:2011da}. 
It's worth noting that the inclination angle of GW170817 was roughly $160^{\circ}$~\citep{LIGOScientific:2017adf,Mooley:2018qfh,Mooley:2022uqa}.

For the 2G detector network HLVKI, the average SNR from the higher modes, before the baseline mode enters the detector band, 
falls below the threshold SNR $\rho_{\rm th}$ in all cases. 
Only in near face-on orientations does the average SNR for BNS and BBH with $e_0=0.4$ meet the threshold.
In contrast, all cases under the 3G detector scenarios not only surpass the SNR threshold but also achieve values on the order of $10^2$  for BNS and BBH.
This underscores that the 2G detector network lacks the requisite sensitivity to detect higher harmonic modes before the baseline mode enters the detector band. On the other hand, 3G detectors can amass a substantial SNR, enabling the easy detection of these higher modes. 
 
For localization, we discard cases where the average localization exceeds $10^4~\rm deg^2$. 
In the HLVKI scenario, as expected, the early localization from the higher modes is poor and carries limited significance.
However, the landscape changes dramatically with the 3G detectors.
Using a single ET, the higher modes of BNS can secure an average localization that varies from 600 deg$^2$ at $e_0=0.05$
to 20 deg$^2$ at $e_0=0.4$, all before the baseline mode enters the band. 
Notably, in near face-on orientations, the localization is further enhanced, achieving an order of magnitude of $\mathcal{O}(1)~\rm deg^2$ at $e_0=0.4$.

The inclusion of two CE detectors in conjunction with ET forms a network that can significantly enhance the localization of the higher modes.
As depicted in the right panel of Fig.~\ref{fig:snr_loc_pre}, the early localization solely from higher modes in nearly all cases is superior to $10^3~\rm deg^2$. 
For BNS, the average localization is 144.8, 34.5, 7.5, and 1.4 deg$^2$ 
for $e_0$ values of 0.05, 0.1, 0.2, and 0.4, respectively. 
In near face-on orientations, these localizations sharpen further to 38.5, 9.3, 2.0, and 0.34 deg$^2$, respectively. 
For NSBH, the average localization spans from $22.4$ to $2477.8~\rm deg^2$
and refines to a range between $4.4$ and $461.4~\rm deg^2$ in near face-on orientations. 
The results for BBH align closely in scale with those of BNS.

\begin{figure*}
\includegraphics[width=0.3\textwidth]{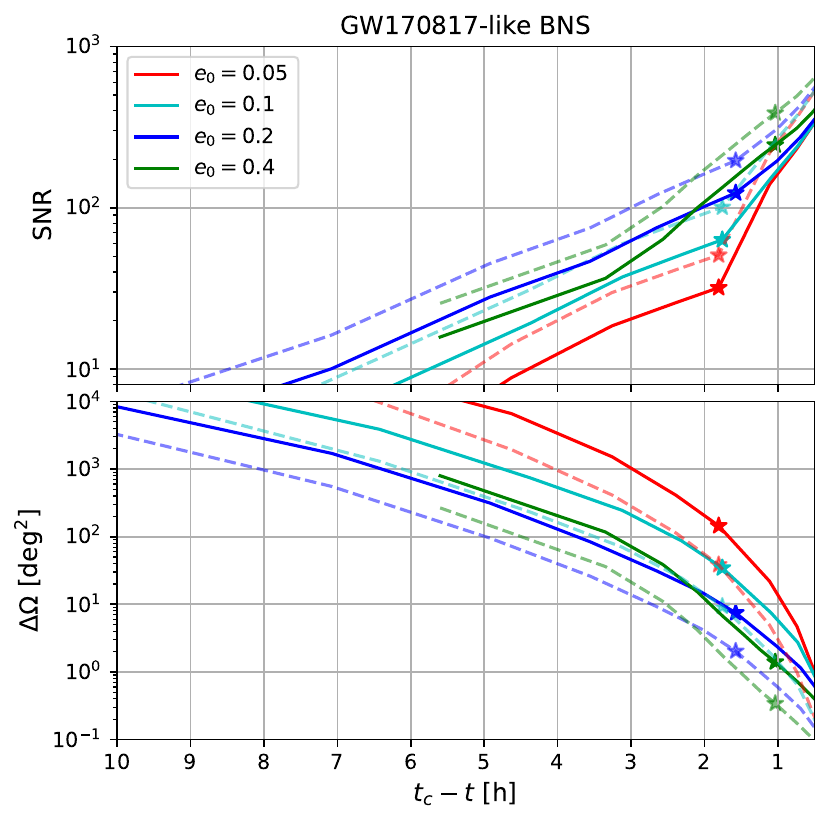}
\includegraphics[width=0.3\textwidth]{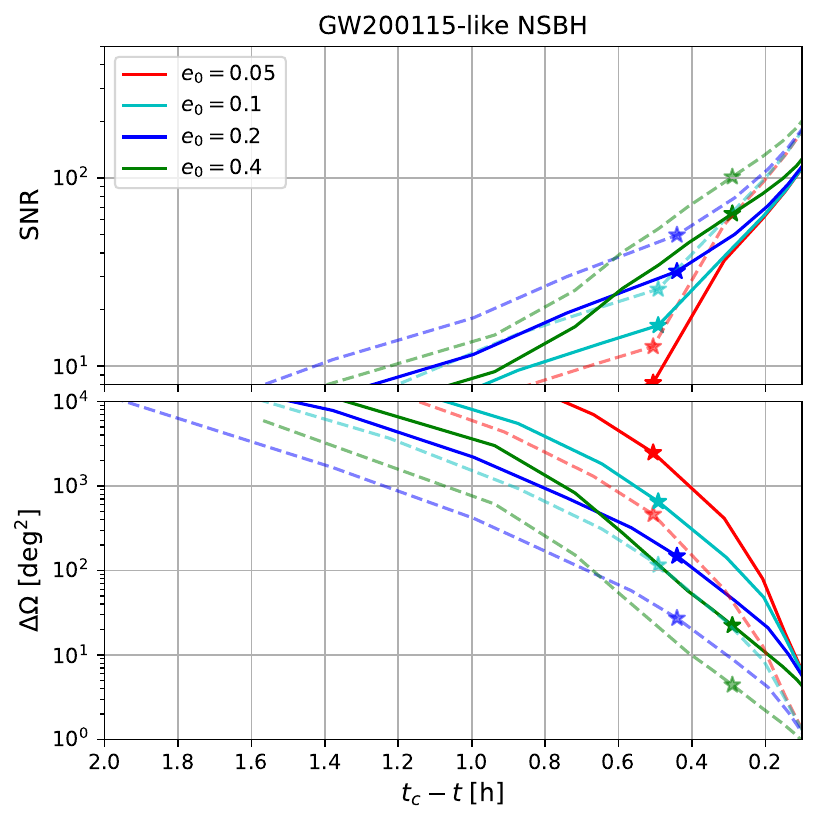}
\includegraphics[width=0.3\textwidth]{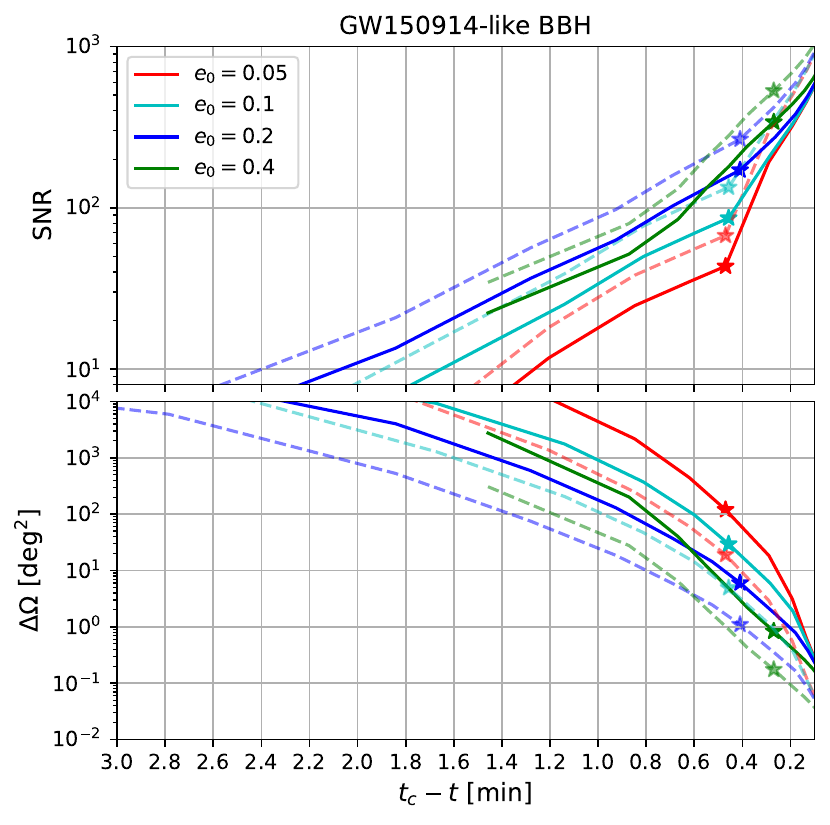}
\caption{The average SNR and localization that can be achieved before the time to merger in the ET+2CE scenario. 
The luminosity distances for BNS, NSBH, and BBH are 40, 290, and 440 Mpc, respectively.
The solid line represents the average of all 1000 $P_{\rm ang}$ samples, 
while the dashed line corresponds to the average with $|\cos\iota|\ge 0.9$ (near face-on orientations). 
Stars indicate $f_2=5$ Hz, marking the point where the dominant baseline mode begins to enter the band.}
\label{fig:snr_loc_evo}
\end{figure*}

To assess how early the higher modes enable us to detect and localize the source, 
we show the evolution of average SNR and localization against time to merger 
for the three typical binaries within the ET+2CE network scenario in Fig.~\ref{fig:snr_loc_evo}.
We calculate the SNR and localization right before $f_2$ reaches 15 distinct values ranging from 1.5 to 50 Hz
(note the baseline mode enters the band at $f_2=5$ Hz, so the results for this particular point just correspond to the right panel of Fig.~\ref{fig:snr_loc_pre}).
We then translate $f_2$ into the time to merger $t_c-t$, to obtain the evolution of SNR and localization against the time to merger.
We place particular emphasis on the period preceding the entry of the baseline mode into the band.
As illustrated in Fig.~\ref{fig:snr_loc_evo}, in general, a larger eccentricity results in a higher SNR and better localization at a given time.
For BNS, with eccentricities ranging from 0.05 to 0.4, the higher modes can achieve the threshold SNR between 4.5 to 7.5 hours prior to the merger. 
This is notably earlier than the entry of the baseline mode into the band.
Localization of $100~\rm deg^2$ can be achieved around 2-4 hours prior to the merger, 
derived solely from the higher modes before the quadrupole mode's entry into the band 
(except for $e_0=0.05$, which is slightly after the baseline mode enters the band).
In the case of NSBH, the threshold SNR can be obtained 0.5-1.3 hours before the merger. 
Solely relying on the higher modes, a localization of $100~\rm deg^2$ can be achieved 30 minutes before the merger when $e_0=0.4$.
The results of BBH are similar to that of BNS, though with a significantly shorter time to merger: 
1.3-2.2 minutes for the threshold SNR and 30-50 seconds for a $100~\rm deg^2$ localization.
In near face-on orientations, the results above can be further improved.

\section{Discussion}
In this letter, we aim at checking the idea that whether the higher harmonic modes of eccentric compact binaries
can make a significant contribution to the SNR and localization of the sources, even before the dominant baseline mode enters the detector band. 
Consequently, these higher modes could offer the earliest alert and localization, 
thereby optimizing preparation time for both pre- and post-merger EM counterpart observations.
Our findings hold significance not only for the EM counterpart observations of BNS and NSBH 
but also for the potential EM follow-ups of BBH~\citep{Connaughton:2016umz,Loeb:2016fzn,Mink:2017npg}.

We fixed the distances of the three typical binaries to match the median values of the true events. 
However, our methodology can easily be extended to other distances, 
as there is a relationship in our calculation of SNR and localization: $\rho \sim 1/d_L$ and $\Delta\Omega \sim 1/\rho^2 \sim d_L^2$.



The rate of eccentric binaries remains a topic of active research~\citep{Ye:2019xvf,Samsing:2013kua,Samsing:2017xmd,Samsing:2017rat}. 
These estimates crucially depend on the unknown astrophysics of these classes of compact binaries. 
For BNS, which are the most promising and sought-after early warning candidates, 
the rate of eccentric events could be rare \citep{Ye:2019xvf}. 
However, the detection of even a single such event would be highly significant. 
The methodology proposed in our paper could provide the maximum preparation time for observations, 
thereby significantly increasing the likelihood of successful joint GW and EM observations. 
Given the extreme rarity of EM counterparts in current GW detections, 
a very early warning of a potential GW + EM signal could be of great interest to both the GW community and EM follow-up teams. 
Our approach offers a promising avenue to enhance the detection and understanding of these rare and valuable events.



\begin{acknowledgments}
This work is supported by the National Research Foundation of Korea 2021R1A2C2012473 and 2021M3F7A1082056.
T. Y. is supported by ``the Fundamental Research Funds for the Central Universities'' under the reference No. 2042024FG0009.
R.G.C is supported by the National Key Research and Development Program of China Grant No. 2020YFC2201502 and 2021YFA0718304 and by National Natural Science Foundation of China Grants No. 11821505, No. 11991052, No. 11947302, No. 12235019.
Z.C is supported in part by the National Key Research and Development Program of China Grant No. 2021YFC2203001, in part by the NSFC (No.~11920101003 and No.~12021003), in part by ``the Interdiscipline Research Funds of Beijing Normal University'' and in part by CAS Project for Young Scientists in Basic Research YSBR-006.
\end{acknowledgments}

\bibliography{ref}

\end{document}